\documentclass[useAMS,usenatbib]{mn2e}

\usepackage{graphicx}

\def\aV{\mbox{$\rm A_V$}}
\def\mMV{\mbox{$(m-M)_V$}}
\def\mMo{\mbox{$(m-M)_O$}}
\def\ebv{\mbox{$E(B-V)$}}
\def\evi{\mbox{$E(V-I)$}}

\def\evi{\mbox{$E(V-I)$}}
\def\rc{\mbox{$R_{\rm c}$}}
\def\rl{\mbox{$R_{\rm RDP}$}}
\def\ms{\mbox{$M_\odot$}}
\def\ds{\mbox{$d_\odot$}}
\def\rs{\mbox{$R_\odot$}}
\def\dgc{\mbox{$R_{\rm GC}$}}

\def\tdis{\mbox{$t_{\rm dis}$}}

\title[The OCs Teutsch\,145 and 146]{Open cluster survival within the solar circle: 
Teutsch\,145 and Teutsch\,146\thanks{Based on observations made with the Italian 
Telescopio Nazionale GALILEO (TNG) operated on the island of La Palma, by the 
Fundaci\'on Galileo Galilei of INAF (Istituto Nazionale di Astrofisica) at the 
Spanish Observatorio del Roque de los Muchachos of the Instituto de Astrofisica de 
Canarias}}

\author[C. Bonatto et al.]{C. Bonatto$^1$, S. Ortolani$^2$, B. Barbuy$^3$ and 
E. Bica$^1$\\
$^1$ Departamento de Astronomia, Universidade Federal do Rio Grande do Sul,
Av. Bento Gon\c{c}alves 9500, Porto Alegre 91501-970, RS, Brazil\\
$^2$  Dipartimento di Astronomia, Vicolo dell'Osservatorio 
5, I-35122 Padova, Italy\\
$^3$ Departamento de Astronomia, Universidade de S\~ao Paulo, Rua do Mat\~ao 1226, 
S\~ao Paulo 05508-900, SP, Brazil }

\begin{document}

\maketitle

\begin{abstract}
Teutsch\,145 and Teutsch\,146 are shown to be open clusters (OCs) orbiting well inside 
the Solar circle, a region where several dynamical processes combine to disrupt most OCs 
on a time-scale of a few $10^8$\,yrs. BVI photometry from the GALILEO telescope is used 
to investigate the nature and derive the fundamental and structural parameters of the 
optically faint and poorly-known OCs Teutsch\,145 and 146. These parameters are computed 
by means of field-star decontaminated colour-magnitude diagrams (CMDs) and stellar radial 
density profiles (RDPs). Cluster mass estimates are made based on the intrinsic mass 
functions (MFs). We derive the ages $200^{+100}_{-50}$\,Myr and $400\pm100$\,Myr, and the 
distances from the Sun $\ds=2.7\pm0.3$\,kpc and $\ds=3.8\pm0.2$\,kpc, respectively for 
Teutsch\,145 and 146. Their integrated apparent and absolute magnitudes are $m_V\approx12.4$,
$m_V\approx13.3$, $M_V\approx-5.6$ and $M_V\approx-5.3$. The MFs (detected for stars with
$m\ga1\,\ms$) have slopes similar to Salpeter's IMF. Extrapolated to the H-burning limit,
the MFs would produce total stellar masses of $\sim1400\,\ms$, typical of relatively 
massive OCs. Both OCs are located deep into the inner Galaxy and close to the 
Crux-Scutum arm. Since cluster-disruption processes are important, their primordial 
masses must have been higher than the present-day values. The conspicuous stellar density 
excess observed in the innermost bin of both RDPs might reflect the dynamical effects 
induced by a few $10^8$\,yrs of external tidal stress.
\end{abstract}


\section{Introduction}
\label{Intro}

Regions interior to the Solar circle represent a harsh environment to the long-term 
survival of open clusters (OCs). The low-mass ones in particular, dissolve into the
field in less than $\approx1$\,Gyr (e.g. \citealt{Friel95}; \citealt{OldOCs}). 

Theoretical and N-body predictions (e.g. \citealt{Spitzer58}; \citealt{LG06}; \citealt{BM03}; 
\citealt{GoBa06}; \citealt{Khalisi07}), coupled to observational evidence (e.g. \citealt{vdB57};
\citealt{Oort58}; \citealt{vHoerner58}; \citealt{Piskunov07}) consistently indicate that
the disruption-time scale (\tdis) near the Solar circle is shorter than $\sim1$\,Gyr and
depends on cluster mass as $\tdis\sim M^{0.62}$ (\citealt{LG06}). Thus, $75\la\tdis(Myr)\la300$
should be expected for clusters with an initial mass within $10^2 - 10^3\ms$. Besides, 
disruption processes are more effective for the more centrally located and lower-mass  
OCs (see \citealt{OldOCs} for a review on these effects). 

It is in this context that the discovery and characterisation of new OCs towards the
inner Galactic regions play an important role. Here we establish the nature and 
derive astrophysical parameters of the poorly-studied, faint OCs Teutsch\,145 and 146. 
Both clusters were discovered by Phillip Teutsch in a systematic survey of several Milky 
Way fields near the Galactic plane using red, blue, and infrared First and Second Generation 
DSS images downloaded from the ESO Online Digitized Sky Survey (\citealt{KTA06}). We are 
dealing with $1^{st}$ Galactic Quadrant clusters, which makes them particularly suitable to explore 
dynamical cluster properties within the Solar circle.

Algorithms designed to deal with field-star contamination in densely populated fields, together 
with other tools to study colour-magnitude diagrams (CMDs) and stellar radial density profiles 
(RDPs) have been developed by our group in previous 2MASS\footnote{The Two Micron All Sky Survey, All
Sky data release (\citealt{2mass1997}) - {\em http://www.ipac.caltech.edu/2mass/releases/allsky/}}
studies (e.g. \citealt{ProbFSR}; \citealt{AntiC}; \citealt{LKstuff}). In the near future,
VISTA\footnote{http://www.vista.ac.uk/ } together with other surveys with large telescopes, will 
deepen by about 4 magnitudes the presently-available near infrared photometry in a large area 
throughout the Galactic plane. The new - and deeper - photometry will probably require specific
algorithms to be analysed. Thus, besides the more direct goal of deriving parameters of two 
interesting objects, in the present work we apply our analytical tools to the optical data of 
the two faint open clusters, Teutsch\,145 and 146, obtained with the 3.58m GALILEO telescope
(TNG)\footnote{http://www.tng.iac.es/ }. 

Since they are located in the $1^{st}$ Quadrant (with the associated enhanced 
disruption rates), the heavy field contamination should be properly taken into account
for the intrinsic properties to be assessed. In this context, our main goal in this
work is to determine whether such clusters can be characterised as typical OCs or if 
they present signs of dissolution. In addition, we will derive their fundamental and 
structural parameters, most of these for the first time. 

\begin{table}
\caption[]{Position and angular size}
\label{tab1}
\renewcommand{\tabcolsep}{2.2mm}
\renewcommand{\arraystretch}{1.25}
\begin{tabular}{cccccc}
\hline\hline
Cluster&$\alpha(2000)$&$\delta(2000)$&$\ell$&$b$&D\\
       &(hms)&($\degr\,\arcmin\,\arcsec$)&(\degr)&(\degr)&(\arcmin)\\
(1)&(2)&(3)&(4)&(5)&(6)\\
\hline
Teutsch\,145&18:42:29&$-$05:15:12&27.24&$-$0.41&1.9\\
Teutsch\,146&18:51:34&$+$00:11:10&33.11&$+$0.06&1.6\\
\hline
\end{tabular}
\begin{list}{Table Notes.}
\item Col.~6: optical diameter measured in the DSS images.
\end{list}
\end{table}

This paper is organised as follows. In Sect.~\ref{RecAdd} we provide details on the
observations, photometric calibrations and reductions. In Sect.~\ref{CMDs} we build
the colour-magnitude diagrams, discuss the field decontamination and derive the
fundamental parameters. In Sect.~\ref{struc} we derive structural parameters. In 
Sect.~\ref{MF} we estimate cluster mass and build the mass functions. In 
Sect.~\ref{Discus} we discuss the parameters of both OCs. Concluding remarks are 
given in Sect.~\ref{Conclu}.

\section{Observations}
\label{RecAdd}

Teutsch\,145 and 146 were observed in 2008 June with the 3.58m GALILEO telescope 
(TNG) at La Palma, equipped with the Dolores spectrograph focal reducer.
A EEV 4260 CCD detector  with 2048$\times$2048 pixels, of pixel size 13 $\mu$m 
was used. A pixel corresponds to $0.252"$ on the sky, and the full field of the 
camera is $8.6'\times 8.6'$. Calibration and reductions are described in detail 
in \citet{Ortolani09}

For reddening transformations we use the relations $\aV=3.1\ebv$, $\evi=1.25\ebv$, 
$A_I=1.95\ebv$, and $A_B=1.322\aV$, taken from \citet{Schl98}, which are based on the
extinction curves of \citet{Card89} and \citet{Odon94}. We remark that the individual
values of absorption and reddening in front of the clusters are determined from the 
CMD fitting (Sect.~\ref{DFP}).

\begin{figure}
\resizebox{\hsize}{!}{\includegraphics{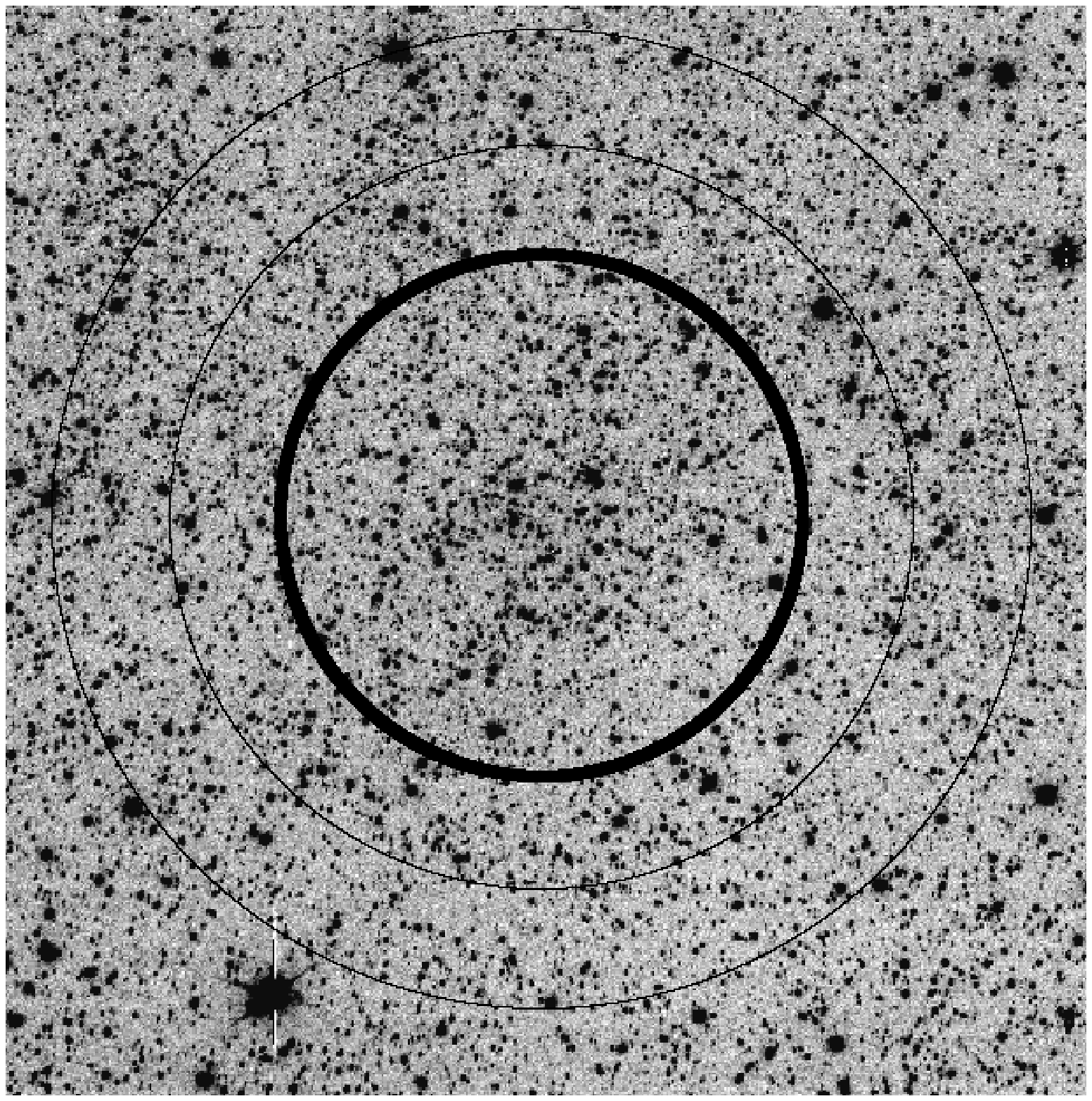}}
\caption[]{Full-frame ($\approx8.5\arcmin\times8.5\arcmin$), 10sec I band image of Teutsch\,145. 
Extractions corresponding to the cluster ($R\approx1.8\arcmin$) and comparison field 
($2.5\arcmin\la R\la3.4\arcmin$) are shown by the inner and outer rings, respectively.}
\label{fig1}
\end{figure}

\begin{figure}
\resizebox{\hsize}{!}{\includegraphics{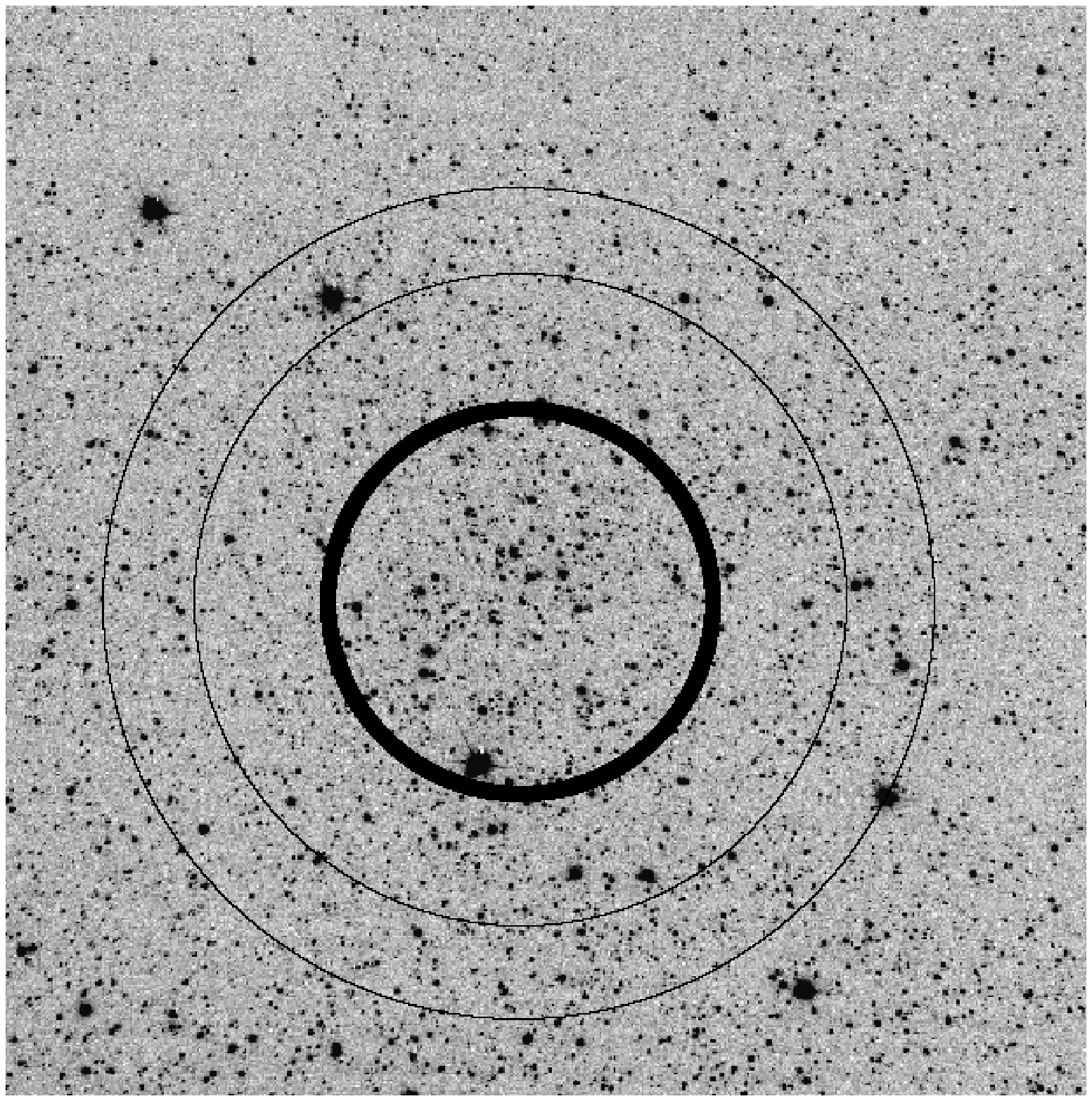}}
\caption[]{Same as Fig.~\ref{fig1} for Teutsch\,146. Cluster and comparison field extractions
are $R\approx1.5\arcmin$ and $2.3\arcmin\la R\la2.9\arcmin$, respectively.}
\label{fig2}
\end{figure}

In Figs.~\ref{fig1} and \ref{fig2} we show 10\,sec I images of both objects and surroundings. 
We also indicate the regions where most of the cluster's stellar content are located, together 
with the respective comparison fields.

\section{Colour-Magnitude Diagrams}
\label{CMDs}

\begin{figure}
\resizebox{\hsize}{!}{\includegraphics{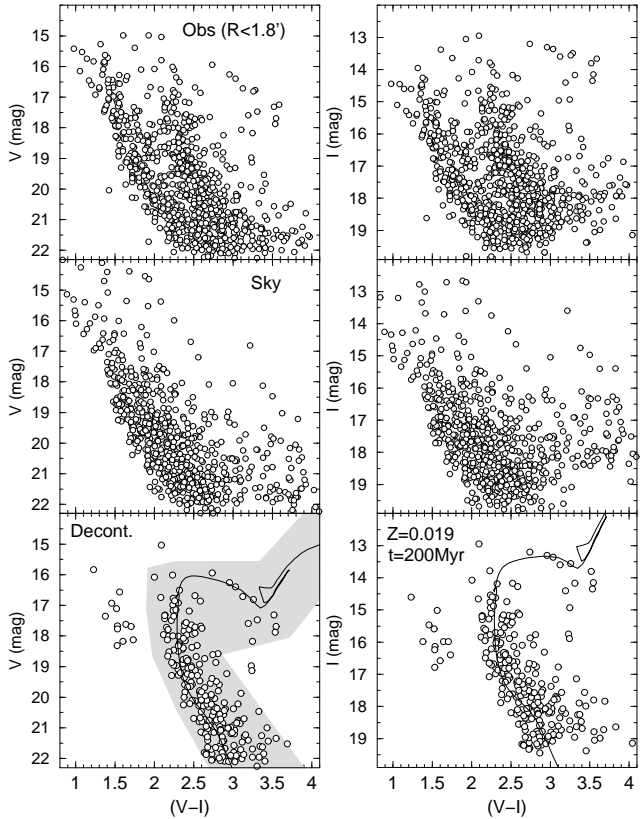}}
\caption[]{$V~vs.~(V-I)$ and $I~vs.~(V-I)$ CMDs of Teutsch\,145. Top panels: observed 
photometry extracted from $R<1.8\arcmin$. Middle: equal-area comparison field extracted 
from $2.9\arcmin\la R\la3.4\arcmin$. Bottom: decontaminated CMDs, including the 
{\em best-fit} isochrone solution. The colour-magnitude filter (shaded polygon) is shown 
in the bottom-left panel.}
\label{fig3}
\end{figure}

\begin{figure}
\resizebox{\hsize}{!}{\includegraphics{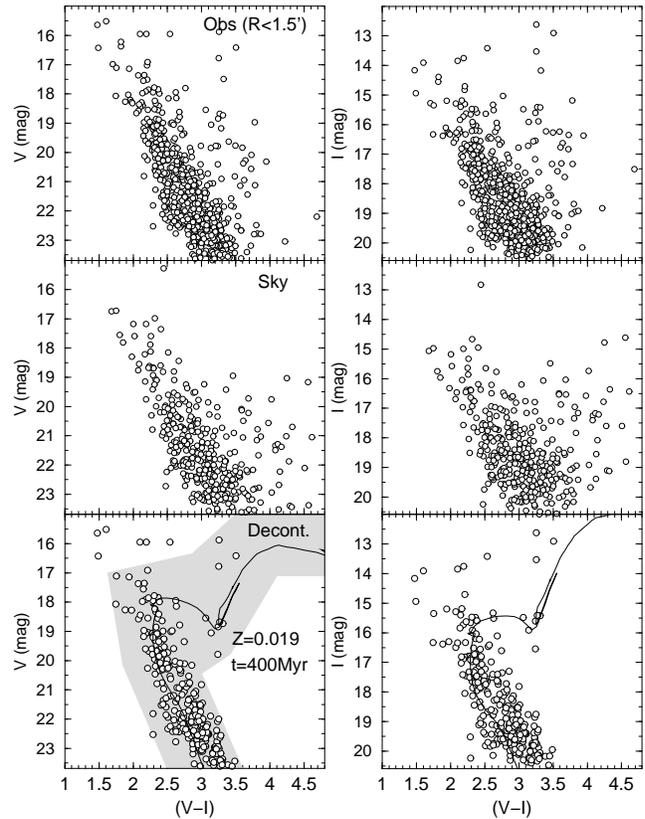}}
\caption[]{Similar to Fig.~\ref{fig3} for the $R<1.5\arcmin$ extraction of Teutsch\,146.
Comparison field extracted from $2.5\arcmin\la R\la2.9\arcmin$.}
\label{fig4}
\end{figure}

CMDs involving V and I of Teutsch\,145 and 146 are shown in Figs.~\ref{fig3} and
\ref{fig4}, respectively. Based on the structural analysis (Sect.~\ref{struc}),
we consider in the top panels the extraction that contains most of each cluster's
stars. When compared with the CMDs extracted from the equal-area comparison field
(middle panels), features typical of evolved OCs emerge from the rather heavy disk and 
bulge contamination: a relatively tight and well-populated main sequence (MS) together 
with a few giant stars. 

\begin{figure}
\resizebox{\hsize}{!}{\includegraphics{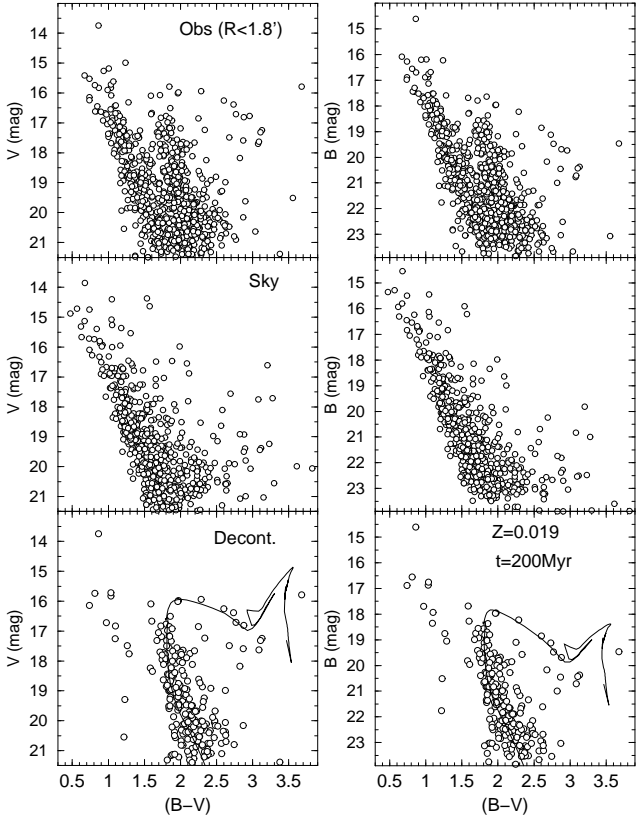}}
\caption[]{Same as Fig.~\ref{fig3} for the $V~vs.~(B-V)$ and $B~vs.~(B-V)$
CMDs of Teutsch\,145.}
\label{fig5}
\end{figure}

CMDs built with the additional B photometry of Teutsch\,145 (Fig.~\ref{fig5}) consistently 
present the same morphology as that implied by Fig.~\ref{fig3}. However, the important 
stellar contamination (middle panels) has to be subtracted before proceeding to a more 
objective interpretation of the CMDs.

\subsection{Field decontamination}

As illustrated by the CMDs of Teutsch\,145 and 146 (Figs.~\ref{fig3}-\ref{fig5}), 
field-stars are an important component in CMDs of clusters projected onto rich fields, 
especially near the disk and bulge. In this paper we work with an algorithm based on 
the three-dimensional decontamination routine designed for the wide-field 2MASS 
photometry (\citealt{BB07}; \citealt{ProbFSR}; \citealt{F1603}). The original algorithm 
was adapted to deal with
photometry obtained with a large telescope and a single colour. For clarity, we recall
the basic procedures. The algorithm divides the magnitude and colour ranges into 
a grid of CMD cells. For a given cluster extraction and comparison field, it estimates 
the relative number densities of probable field and cluster stars present in each cell. 
The estimated number of field stars is subsequently subtracted from each cell. Reference 
cell dimensions are $\rm\Delta\,mag=1.0$ and $\rm\Delta\,colour=0.2$. In addition, we 
minimise spurious results by means of several runs of the decontamination procedure, 
with different input parameters. Here, different cell sizes are considered, with 
$\rm\Delta\,mag$ and $\rm\Delta\,colour$ taken as 0.5, 1.0 and 2.0 times the reference 
values. Also, the cell grid is shifted by -1/3, 0 and +1/3 of the respective cell
size in both the colour and magnitude axes. Taking together all the grid/cell size 
setups, we are left with 81 different and independent decontamination combinations. 
Stars are ranked according to the number of times they survive each run. Finally, only 
the highest ranked stars are considered as cluster members and transposed to the 
respective decontaminated CMD.

Since the GALILEO field is somewhat limited, covering about $8.5\arcmin\times8.5\arcmin$,
we take as comparison field the rings within $2.5\arcmin\la R\la3.4\arcmin$ and 
$2.3\arcmin\la R\la2.9\arcmin$, respectively for Teutsch\,145 and 146. This geometrical
setup (Figs.~\ref{fig1} and \ref{fig2}) prevents border effects and minimises the 
oversubtraction of member stars at the cluster's outskirts. Indeed, the number density 
of stars in the comparison fields corresponds to about 1/4 that in the central parts
(Fig.~\ref{fig7}).

The decontaminated CMDs are shown in the bottom panels of Figs.~\ref{fig3} to 
\ref{fig5}. As expected, essentially all contamination is removed, leaving stellar 
sequences typical of reddened and evolved OCs. The decontaminated CMDs also show
some scatter that, in the bright CMD sequences, may occur from low-number statistics 
and, consequently decontamination inefficiency (this issue is thoroughly discussed in 
\citealt{BB07}). However, much of the scatter among giants is due to binarism, as 
detected from proper motions and radial velocity variations (e.g. \citealt{Hole09}). 
In the blue sequences, blue stragglers and binarism in general are important sources 
of scatter (e.g. \citealt{Geller09}). As discussed in \citet{BB07}, differential 
reddening is also a potential source of reddening. However, since {\em (i)} the sampled 
regions are relatively small ($\la3.5\arcmin$), {\em (ii)} the foreground absorption 
is moderate ($\aV\la6$\,mag) in both cases (Sect.~\ref{DFP}), {\em (iii)} the distribution
of stars in the I images (Figs.~\ref{fig1} and \ref{fig2}) is rather uniform and, {\em (iv)}
the cell dimensions used in the decontamination algorithm are wide enough to minimise 
differential reddening effects, the differential reddening is not expected to be a major 
source of scatter in the CMDs of Teutsch\,145 and 146. The similar decontaminated CMD 
morphologies indicate comparable ages for both objects.

\subsection{Fundamental parameters}
\label{DFP}

We base the fundamental parameter derivation on the field-decontaminated CMD morphologies 
(Figs.~\ref{fig3}-\ref{fig5}), using as a constraint the different combinations of magnitudes 
and colours. Fits with Padova isochrones (\citealt{Girardi2002}) are made {\em by eye},
taking the MS and giant stars as constraint. The adopted results are shown in Figs.~\ref{fig3} 
to \ref{fig5} (bottom panels) and discussed below.

\begin{figure}
\resizebox{\hsize}{!}{\includegraphics{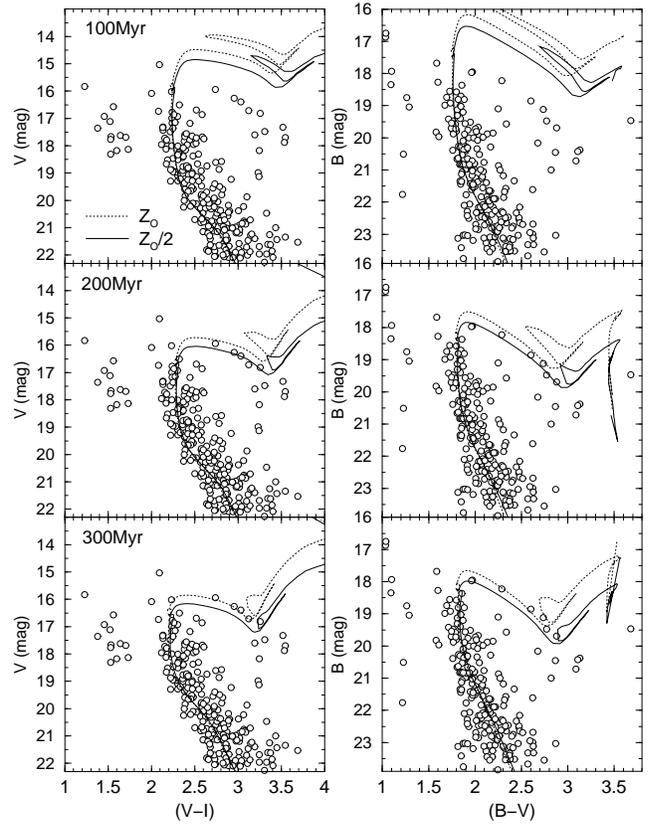}}
\caption[]{Age and metallicity determination for Teutsch\,145.}
\label{fig6}
\end{figure}

{\tt Teutsch\,145:}
At first sight, the decontaminated features denote an OC a few $10^8$\,yrs old and nearly 
solar metallicity. Thus, we search for solutions with isochrones with the ages 100, 200 
and 300\,Myr, of solar and half solar metallicities. This age/metallicity search is 
illustrated in Fig.~\ref{fig6}, where we use the decontaminated $R=1.8\arcmin$ extraction 
and the $V~vs.~(V-I)$ and $B~vs.~(B-V)$ CMDs. We require that all solutions provide
a similar representation of the MS, the more statistically significant CMD feature.

Clearly, Teutsch\,145 is older than 100\,Myr (top panels) but not much older than 300\,Myr 
(bottom panels). Besides, the solar metallicity isochrone appears to produce the best fit 
for the best age solution, 200\,Myr (middle panels). Thus, taking into account the above 
range of ages and metallicities, we found that the best solution corresponds to the age 
$200^{+100}_{-50}$\,Myr and solar metallicity, although lower metallicity values cannot 
be ruled out. 

With the adopted (200\,Myr) isochrone solution, the fundamental parameters of Teutsch\,145 
are the reddening $\evi=2.37\pm0.02$, which converts (Sect.~\ref{RecAdd}) to $\ebv=1.90\pm0.02$, 
a total-to-selective absorption $\aV=5.88\pm0.06$, the observed and absolute distance moduli
$\mMV=18.0\pm0.2$ and $\mMo=12.12\pm0.21$, respectively, and the distance from the Sun 
$\ds=2.7\pm0.3$\,kpc. We adopt $\rs=7.2\pm0.3$\,kpc (\citealt{GCProp}) as the Sun's distance to 
the Galactic centre to compute Galactocentric distances\footnote{Derived by means of the Globular
Cluster (GC) spatial distribution. Recently, \citet{Trippe08} found $\dgc=8.07\pm0.32$\,kpc while 
\citet{Ghez08} found $\dgc=8.0\pm0.6$\,kpc or $\dgc=8.4\pm0.4$\,kpc, under different assumptions.}.
For $\rs=7.2$\,kpc, the Galactocentric distance of Teutsch\,145 is $\dgc=5.0\pm0.2$\,kpc, which puts 
it $\approx2.2$\,kpc inside the Solar circle. This solution is shown in Figs.~\ref{fig3} and
\ref{fig5}. 

The above fundamental parameters are used to compute integrated magnitudes and colours, 
for the stars within $R=1.8\arcmin$ isolated by the colour-magnitude filter 
(Sect.~\ref{struc}). The integrated apparent magnitudes are $m_V\approx12.4$ and $m_I\approx9.8$, 
the reddening-corrected colour is $(V-I)=+0.40\pm0.03$, and the absolute magnitudes are 
$M_V=-5.6\pm0.1$ and $M_I=-6.0\pm0.1$. Compared to the set of OCs analysed by \citet{Lata02},
Teutsch\,145 is intrinsically somewhat brighter than the mean value of the distribution. 

{\tt Teutsch\,146:}
A similar approach is applied to derive the fundamental parameters of Teutsch\,146.  
Given the similarities with the CMD of Teutsch\,145, uncertainties in age and metallicity
of the same order are expected. Indeed, its age is within $400\pm100$\,Myr, biased to
the solar metallicity (Fig.~\ref{fig4}).

The fundamental parameters computed for the 400\,Myr solution are: $\evi=2.29\pm0.01$ 
($\ebv=1.83\pm0.01$ or $\aV=5.68\pm0.02$), $\mMV=18.6\pm0.1$, $\mMo=12.92\pm0.10$, 
$\ds=3.8\pm0.2$\,kpc, and $\dgc=4.5\pm0.1$\,kpc, thus $\approx2.7$\,kpc inside the 
Solar circle. Also, $m_V\approx13.3$, $m_I\approx10.7$, $(V-I)=+0.44\pm0.03$, 
$M_V=-5.3\pm0.1$ and $M_I=-5.8\pm0.1$. Similarly to Teutsch\,145, Teutsch\,146 
can also be considered as an intrinsically bright OC.

\section{Cluster structure}
\label{struc}

We use the RDPs, defined as the projected stellar number density around the cluster 
centre to derive structural parameters. Stars with colours unlike those of the cluster 
CMD morphology are excluded by means of the colour-magnitude filters (shown in 
Figs.~\ref{fig3} and \ref{fig4}). This procedure enhances the RDP contrast relative to 
the background, especially in crowded fields (e.g. \citealt{BB07}; \citet{OldOCs}; 
\citet{AntiC}). 

\begin{figure}
\resizebox{\hsize}{!}{\includegraphics{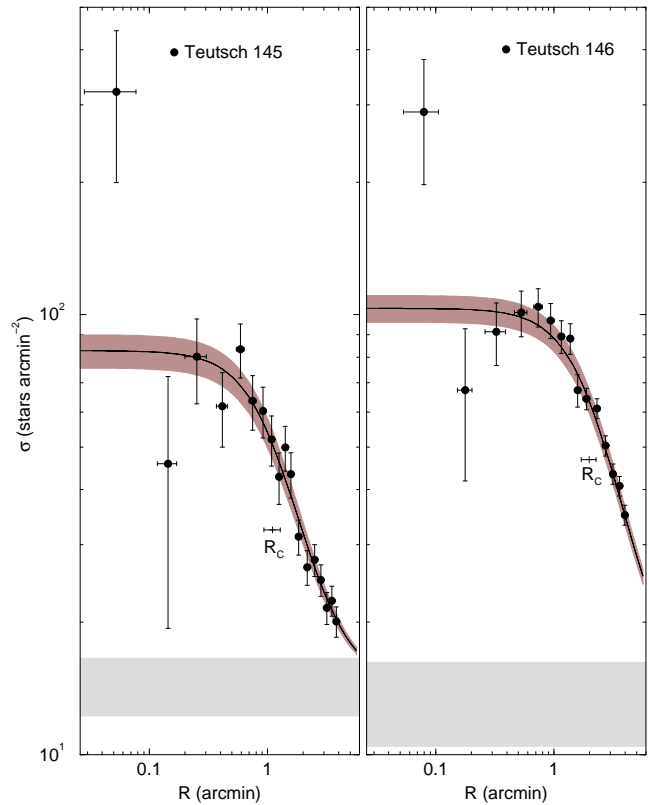}}
\caption[]{Stellar RDPs built with colour-magnitude filtered photometry together with
the best-fit King-like profile (solid line), the $1\sigma$ uncertainty and the background 
level (shaded polygon). Both OCs present a cusp in the innermost RDP bin.}
\label{fig7}
\end{figure}

Rings of increasing width with distance from the cluster centre are used to preserve
spatial resolution near the centre and minimise noise at large radii. The $R$ coordinate 
(and uncertainty) of each ring corresponds to the average position and standard deviation 
of the stars inside the ring. The RDPs of Teutsch\,145 and 146 are shown in Fig.~\ref{fig7}.
The effective (i.e., avoiding border effects) radial range of both RDPs reaches about 
$4\arcmin$, which is clearly less than the cluster size. 


\begin{table*}
\caption[]{Derived structural parameters}
\label{tab2}
\renewcommand{\tabcolsep}{3.7mm}
\renewcommand{\arraystretch}{1.25}
\begin{tabular}{cccccccccc}
\hline\hline
Cluster&$\sigma_{bg}$&$\sigma_0$&\rc&&$1\arcmin$&&$\sigma_{bg}$&$\sigma_0$&\rc\\
       &$\rm(*\,\arcmin^{-2})$&$\rm(*\,\arcmin^{-2})$&(\arcmin)&&(pc)&&
$\rm(*\,pc^{-2})$&$\rm(*\,pc^{-2})$&(pc)\\
(1)&(2)&(3)&(4)&&(5)&&(6)&(7)&(8)\\
\hline
Teutsch\,145&$14.4\pm2.2$&$68.4\pm7.2$&$1.1\pm0.2$&&0.774&&$24.0\pm3.6$&$114.2\pm12.1$&$0.86\pm0.13$\\
   
Teutsch\,146&$13.3\pm6.1$&$90.0\pm7.2$&$2.0\pm0.3$&&1.113&&$10.8\pm5.2$&$72.6\pm5.8$&$2.21\pm0.31$\\
\hline
\end{tabular}
\begin{list}{Table Notes.}
\item Col.~5: arcmin to parsec scale. 
\end{list}
\end{table*}

To derive cluster structural parameters, we fit the RDPs with the analytical function 
$\sigma(R)=\sigma_{bg}+\sigma_0/(1+(R/R_c)^2)$, where $\sigma_{bg}$ is the residual 
background density, $\sigma_0$ is the central density of stars, and \rc\ is the core radius. 
Formally, it is similar to the \cite{King1962} function that describes the surface-brightness 
profiles in the central parts of GCs. However, in the present cases it is applied to star 
counts, with equivalent results (e.g. \citealt{StrucPar}). Given the limited radial range 
of the RDPs, the background level is little constrained and thus, its fit value has 
large error bars. Besides, the actual cluster size cannot be precisely determined. The central
regions, on the other hand, are well sampled. With the above restrictions in mind, the best-fit 
solutions (together with the uncertainties) are shown in Fig.~\ref{fig7}, and the parameters are 
given in Table~\ref{tab2}. 

Within uncertainties, the adopted King-like function describes both RDPs along most of the 
detected radius range. However, the innermost bin ($R\la0.1\arcmin$) presents a significant 
excess over the fit in both cases. Such a cusp has been attributed to a post-core collapse 
structure in old star clusters, like those detected in some GCs (e.g. \citealt{TKD95}). 
Gyr-old OCs, e.g. NGC\,3960 (\citealt{N3960}) and LK\,10 (\citealt{LKstuff}), also present 
this dynamical evolution-related feature. Thus, the presence of such features in clusters a 
few $10^8$\,yrs old located in the inner Galaxy (Sect.~\ref{DFP}) is not unusual.

Based on the extrapolation of the King-like fit into the background (taking into account 
the respective uncertainties), we estimate that the cluster radius of Teutsch\,145 is
about $\rl\approx9\arcmin\pm2\arcmin~(\approx7$\,pc). For Teutsch\,146 we estimate the
somewhat larger value $\rl\approx14\arcmin\pm4\arcmin~(\approx16$\,pc).

Compared to the core radii derived for a sample of relatively nearby OCs by 
\citet{Piskunov07}, the King-like values of the present OCs ($\rm 0.9\la\rc(pc)\la2.2$) 
fall around the mean value of that distribution. However, these values should be taken only 
as representative, since the inner region of both OCs clearly does not follow the King-like 
profile.

\section{Mass estimate}
\label{MF}

About 5 mags of the field-decontaminated MS (together with a few giants) of Teutsch\,145
and 146 are detected by the GALILEO photometry (Figs.~\ref{fig3}-\ref{fig5}), which can 
be used to build the mass function $\left(\phi(m)=\frac{dN}{dm}\right)$ and estimate the 
mass stored in stars. 

The decontamination algorithm excludes stars in integer numbers from the CMDs, and thus, 
it should be used essentially to determine the intrinsic CMD morphology. However, when 
magnitude (or mass) bins are considered, the bin-to-bin subtraction of the comparison 
field contribution (normalised to the same projected areas) is expected to produce 
fractional numbers, which should be taken into account by, e.g. the MFs or the cluster 
mass (e.g. \citealt{DetAnalOCs}). Thus, the following analyses are based on colour-magnitude 
filtered photometry (Figs.~\ref{fig3} and \ref{fig4}).

For the present purposes we consider the regions within $R\le1.8\arcmin$ and 
$R\le1.5\arcmin$ (Figs.~\ref{fig1} and \ref{fig2}), respectively for Teutsch\,145
and 146, which correspond to $\approx1.6$\,pc in both cases. The effective MS stellar 
mass ranges are $1.17\leq m(\ms)\leq4.11$ (Teutsch\,145) and $0.97\leq m(\ms)\leq3.17$.
(Teutsch\,146). The number of member MS and giant stars is derived by counting the stars 
(in bins of $\Delta\,V=0.5$\,mag), and subtracting those in the field (normalised to the 
same area). The corresponding stellar mass in each magnitude bin is taken from the 
mass-luminosity relation derived from the isochrone fits (Sect.~\ref{DFP}). We found 
$m_{MS}=397\pm37\,\ms$ and $m_{giant}=45\pm12\,\ms$, respectively for the MS and giant 
stars of Teutsch\,145, and the similar values $m_{MS}=409\pm35\,\ms$ and 
$m_{giant}=70\pm16\,\ms$ for Teutsch\,146. Thus, the respective total stellar mass values 
inferred within the spatial region considered are $\approx440\,\ms$ and $\approx480\,\ms$. 

\begin{figure}
\resizebox{\hsize}{!}{\includegraphics{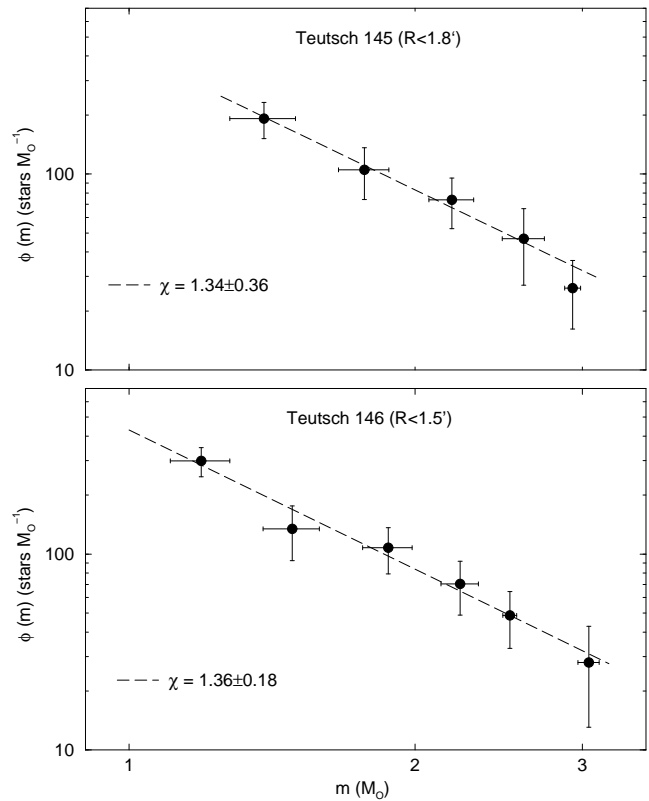}}
\caption[]{The intrinsic main-sequence mass functions (black circles) are fitted by the 
function $\phi(m)\propto m^{-(1+\chi)}$ (dashed line).}
\label{fig8}
\end{figure}

With the above data we build the MF for the MSs (Figs.~\ref{fig8}). Both MFs are well 
represented by the function $\phi(m)\propto m^{-(1+\chi)}$, with the slope $\chi=1.34\pm0.36$
and $\chi=1.36\pm0.18$, respectively for Teutsch\,145 and 146. These values agree with the 
$\chi=1.35$ of \citet{Salpeter55} initial mass function (IMF). This is not a
surprising result, since recent works raise the possibility of a universal initial
mass function that, for the mass range $m\ga1\,\ms$, is essentially Salpeter (e.g.
\citealt{Kroupa2001}). Besides, Salpeter-like slopes also occur (for $m\ga1\,\ms$)
in a variety of OCs younger than about 1\,Gyr (e.g. Fig.~12 in \citealt{Pi5} for the 
near-infrared, and \citealt{MM07} in the optical). 

Finally, we estimate the total stellar mass (within $\approx1.6$\,pc) by extrapolating the 
observed MFs down to the H-burning mass limit ($0.08\,\ms$). We assume the universal IMF of
\citet{Kroupa2001}, which is characterised by the slopes $\chi=0.3\pm0.5$ for the range 
$0.08\leq m(\ms)\leq0.5$ and $\chi=1.3\pm0.3$ for $0.5\leq m(\ms)\leq1.0$. We obtain
$m_{extr}\approx1400\pm500\,\ms$ in both cases, a value somewhat higher than the mean 
cluster mass with respect to the nearby OC distribution of \citet{Piskunov07}. For these 
values, the mass to light ratios are $M/L_V\approx0.10$ and $0.12$, respectively for 
Teutsch\,145 and 146, consistent with their ages (e.g. \citealt{Bica88}).


The above mass values should be taken as lower limits, since the clusters are larger 
than the region ($\approx1.6$\,pc) within which the mass functions were computed. 
Besides, given the restricted spatial range of the observations, the comparison fields 
are located at the outskirts of the clusters themselves, and thus, an oversubtraction 
of member stars (mostly low-mass stars) certainly occurred.

\section{Discussion} 
\label{Discus}

\begin{figure}
\resizebox{\hsize}{!}{\includegraphics{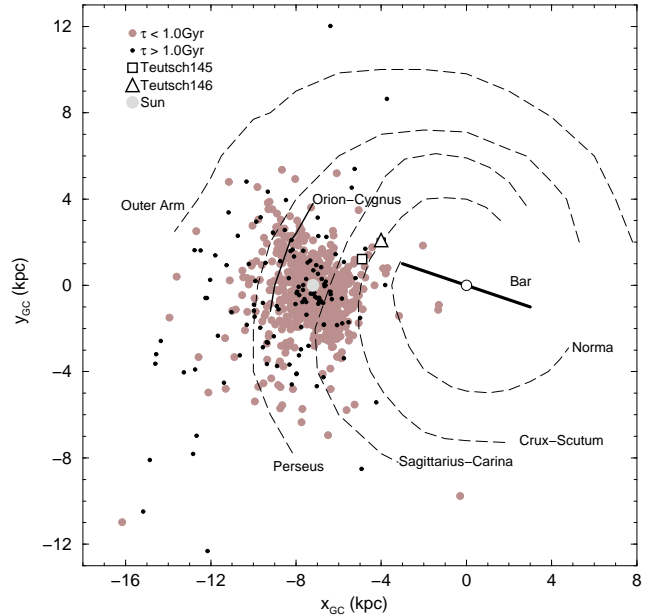}}
\caption{Projected distribution of the present star clusters compared to the WEBDA 
OCs younger (gray circles) and older than 1\,Gyr (black dots). Clusters are shown 
on a schematic projection of the Galaxy, as seen from the North pole, with 7.2\,kpc 
as the Sun's distance to the Galactic centre. Main structures are identified.}
\label{fig9}
\end{figure}

The positions of Teutsch\,145 and 146, projected onto the Galactic plane, are given
in Fig.~\ref{fig9}, which shows the spiral arm structure of the Milky Way based on
\citet{GalStr} and \citet{DrimSper01}, derived from HII regions and molecular clouds 
(e.g. \citealt{Russeil03}). The Galactic bar is shown with an orientation of 14\degr\ 
and 6\,kpc of total length (\citealt{Freuden98}; \citealt{Vallee05}). 

Both OCs are compared with the spatial distribution of the OCs with known age and distance 
from the Sun given in the WEBDA database. Two age groups are considered, clusters younger 
and older than 1\,Gyr. In the inner Galaxy, dynamical interactions with the disk, 
the tidal pull of the Galactic bulge, and collisions with giant molecular clouds, tend 
to destroy OCs, especially the poorly-populated ones, on a timescale of a few $10^8$\,yr 
(e.g. \citealt{BLG01}). In this context, it should be expected to find old OCs preferentially
outside the Solar circle (Fig.~\ref{fig9}), a region with lowered tidal stress from the 
Galaxy and with less probability of
encounters with giant molecular clouds (e.g. \citealt{vdBL80}; \citealt{Friel95};
\citealt{OldOCs}). A similar scenario, with the outer disk hosting predominantly the old
population, has been observed  in other galaxies as well, e.g. NGC\,300 (\citealt{Vlad09}). 
The presence of bright stars in young OCs, on the other hand, allows them to be detected 
farther than the old ones, especially towards the central Galaxy. Central regions more 
distant than $\approx2$\,kpc begin to be critically affected by completeness effects (due 
to crowding and high background levels) and enhanced disruption rates (e.g. 
\citealt{DiskProp}). Besides, all directions show a depletion in the number of OCs 
detected farther than $\approx2$\,kpc. 

Teutsch\,145, and especially 146, are located close to the Crux-Scutum arm, among the
most (centrally) distant OCs so far detected. Since they are projected essentially low on
the disk ($|b|\la0.4\degr$), tidal stresses related to collisions with the spiral arm
may have induced dynamical effects on them (e.g. \citealt{GAP07}; \citealt{AntiC}).
As discussed in Sect.~\ref{struc}, the central cusp in the RDPs may be an example of 
such an effect. The preferential low-mass star loss, and the resulting MF flattening,
might also reflect this mechanism. However, given the distance of the clusters 
(Sect.~\ref{DFP}), the GALILEO photometry could not detect the sub-solar mass range
(Sect.~\ref{MF}). 

In any case, for both OCs to retain a significant amount of stellar mass (at least
$m\sim1400\,\ms$) after spending a few $10^8$\,yrs in the inner Galaxy, their 
primordial masses might have been significantly higher than the present values.

\section{Concluding remarks}
\label{Conclu}

We use BVI photometry obtained with the TNG (3.58m) telescope to derive astrophysical 
parameters and investigate the nature of the two optically faint and poorly known 
OCs Teutsch\,145 ($m_V=12.4$) and Teutsch\,146 ($m_V=13.3$). Located in the $1^{st}$ 
Quadrant, both OCs present heavily field-contaminated CMDs, which makes their nature 
and properties difficult to establish from optical studies in smaller telescopes.

Decontaminated CMDs show that the two clusters exhibit similar properties, basically 
a well-populated MS together with a few giants. From these we derive ages of
$200^{+100}_{-50}$\,Myr and $400\pm100$\,Myr, and distances from the Sun
$\ds=2.7\pm0.3$\,kpc and $\ds=3.8\pm0.2$\,kpc, respectively for Teutsch\,145 and 146.
Their mass functions, detected for stars more massive than $\approx1\,\ms$, present slopes 
similar to Salpeter's IMF. Extrapolated to the H-burning limit, both cluster masses are 
of the order of $1400\,\ms$, which would characterise them as relatively massive OCs. 
Intrinsically, they are bright OCs, with integrated $M_V\approx-5.6$ and $M_V\approx-5.3$, 
respectively. However, given the limited spatial range of the observations, the present-day 
mass values may be somewhat higher.

Teutsch\,145 and 146 are located in the inner Galaxy (more than 2\,kpc inside the Solar 
circle), a region where cluster-disruption processes are important. Besides, they are close 
to the Crux-Scutum arm. Thus, given the ages, their primordial masses must have been higher 
than the present-day values. With respect to the radial density distribution of stars, they 
both present a cusp in the innermost region, which might reflect dynamical effects induced 
by the important external tidal stresses acting along a few $10^8$\,yrs. The present
analysis may shed light on issues such as cluster stability, tidal disruption rates and 
the future cluster evolution in such harsh environment.


\section*{Acknowledgements}
We thank an anonymous referee for interesting suggestions.
We acknowledge partial financial support from CNPq (Brazil) and the Ministero 
dell'Universit\`a e della Ricerca Scientifica e Tecnologica (MURST), Italy. 
This research has made use of the WEBDA database, operated at the Institute 
for Astronomy of the University of Vienna.

\label{lastpage}
\end{document}